# Climatology of Mars Topside Ionosphere during Solar Cycles 24 and 25 using MAVEN Dataset of 2015-2024

Lot Ram[1,*], Chanchal Singh[1], Diptiranjan Rout[2], Aadarsh Raj Sharma[1], Sumanta Sarkhel[1,†,*]

**Abstract**

The Mars ambient space environment evolves with the varying solar activity. Understanding the Martian space environment, particularly the topside ionosphere across different phases of Solar Cycles (SC) 24 & 25 remains a key research gap in planetary ionospheric science. In this study, we utilized the NASA's Mars Atmosphere and Volatile EvolutioN (MAVEN) mission data (150-500 km) from Martian years 32-38 (2015-2024) during solar quiet-time. This study investigated the behavior of topside ionosphere (e$^-$, $CO_2^+$, $O_2^+$, $NO^+$, $OH^+$, $O^+$, $N^+$ & $C^+$) across different phases of SC over the northern hemisphere. A significant variation in ionosphere is observed over low-latitude (0-30°N) with higher densities compared to mid-latitude (31-60°N) across SC. Additionally, we found that the Martian northern ionospheric densities were highest during solar maximum phase on both dayside and nightside compared to low active phases. The dayside densities were approximately 1-2 orders higher compared to those on the nightside. The electron and molecular ions densities increased by factors of 1-5 and 1-13, respectively. While $O^+$ ion density was enhanced by nearly 2-2.5 times, along with an upliftment of ~40-50 km in the peak height. The enhanced dayside densities are attributed to the elevated solar irradiance (~1.4-2 times) and varying solar wind flux. Furthermore, the enhanced day-to-night plasma transport and elevated solar electron flux during maxima, higher by ~33-66% than during low-activity, can contribute to the increased nightside ionization. This work, for the first time, uses long-term MAVEN datasets across the descending-to-maxima phases of SC to reveal climatology of Martian topside ionosphere.

[*]Corresponding authors: sarkhel@ph.iitr.ac.in; l_ram@ph.iitr.ac.in
[1]Department of Physics, Indian Institute of Technology Roorkee, Roorkee-247667, India
[2]National Atmospheric Research Laboratory, Gadanki, 517112, India
[†]Centre for Space Science and Technology, Indian Institute of Technology Roorkee, Roorkee-247667, India

# 1. Introduction

The planetary ambient space environments, such as the Earth and Mars, evolve with respect to the varying solar wind particle flux and electromagnetic radiation. Their axial-tilt causes seasons; in particular, Mars' elliptical orbit and varying solar activity leads to annual variations in solar irradiance, resulting in unequal heating across atmosphere-to-surface. Currently, the NASA's Mars Atmosphere and Volatile EvolutioN (MAVEN) mission is exploring the Mars ambient space, including the ionosphere at various latitudes, longitudes, local time, solar zenith angles (SZA), and altitudes. The ionosphere of Mars is a weakly ionized region starting from 100 km and extending to nearly 500 km altitude (K. Peter et al. 2024; P. Withers 2009). The Martian ionosphere is primarily formed through the photoionization of neutral atmospheric species—mainly $CO_2$, followed by O and CO (J. L. Fox & A. Dalgarno, 1979; K. Hensley & P. Withers, 2022). Specifically, solar extreme ultraviolet (EUV) photons ionize $CO_2$ to produce $CO_2^+$, which rapidly undergoes a charge exchange reaction with neutral O to form $O_2^+$—the dominant ion in the Martian ionosphere. The $O_2^+$ ion eventually loses through the dissociative recombination with electrons at a slower rate (R. Schunk & A. Nagy 2007; P. Withers 2009). The Martian ionosphere comprises the M1, M2, and topside regions. The M2 layer (~130–160 km) and M1 layer (~105–120 km) correspond to primary and secondary ionization peaks, primarily formed and controlled by extreme ultraviolet (EUV) and soft X-ray (SXR) radiation, respectively (P. A. Cloutier et al. 1969; J. L. Fox & K. E. Yeager, 2006; P. Withers, 2009; S. W. Bougher et al. 2017). These regions are dominated by the photochemical processes. The plasma region existing above the peak (specifically above 180 km) and extending up to 400-500 km is collectively called as topside ionosphere, where, the transport processes can affect the ionospheric densities through diffusion and various escape mechanisms (H. Rishbeth et al. 1963; J. K. Fox 1997; V. A. Krasnopolsky 2002; J. Y. Chaufray et al. 2014; M. Mendillo et al. 2011). The topside ionosphere is directly exposed to the incoming solar radiation and particles and acts as a reservoir that supplies plasma to atmospheric escape while interacting with energetic solar wind and radiation (for e.g., J. L. Fox 2009; B. M. Jakosky et al. 2015b; Y. J. Ma & A. F. Nagy 2017). To decipher the Mars climatic evolution over a long time-scales, the understanding of the topside ionosphere is crucial, which is the primary focus of this study.

Previous work have explored a variable plasma densities across seasons (Z. Girazian et al. 2023; F. González-Galindo et al. 2021; N. Yoshida et al.

2021) and solar activities (Z. Girazian et al. 2023; K. Hensley & P. Withers 2022), including the effects during different space weather conditions (M. Mendillo et al. 2006; N. J. T. Edberg et al. 2010; B. M. Jakosky et al. 2015b; S. V. Thampi et al. 2018; ; E. M. Theimann et al. 2018; L. Ram et al. 2023, 2024a,b,c; A. R. Sharma et al. 2025; L. Liu et al. 2025), and dust storms (X. Fang et al. 2019; M. Felici et al. 2020). The study by B. Sánchez-Cano et al. (2018) has demonstrated that the seasonal variation of total electron content (TEC; between 100 and 200 km) is closely related to the varying solar irradiance. The study indicates an important role of atmospheric cycles ($CO_2$ and Water Cycles) in the modulation of the thermospheric region that eventually affects the ionosphere, specifically during the northern winter. Utilizing the radio-occultation measurements from the Mars Global Surveyor, K. Hensley & P. Withers (2021) reported a modest change in the electron densities at and just above the peak at SZA between 76° and 78° with respect to different solar active periods. They observed that changes in electron densities throughout the photochemical region (100-200 km) are linked with the variation in neutral atmosphere due to varying solar irradiance. Further, the work by N. Yoshida et al. (2021) found a seasonal sinusoidal variation in the ionospheric species ($CO_2^+$ and $O_2^+$) at 200 km altitude, which is directly linked to the neutral $CO_2$ variation along the thermosphere-ionosphere region (100-200 km). Although, they found a less significant variation for $O^+$ density with respect to the seasons. In addition, K. Hensley, P. Withers & E. M. Theimann (2022) reported a significant enhancement in the ion and neutral densities (SZA < 25°) between 120 to 200 km altitudes, during the high (17–22 April 2015) solar irradiance period compared to the low (20–23 October 2017) solar irradiance period. Recently, V. Mukundan et al. (2025) have demonstrated a 40% decrease in plasma density from mid-to-low solar activity, along with a 5-10% variability attributed to atmospheric expansion/contraction, specifically during the declining phase of Solar Cycle (SC) 24.

Most previous studies targeted the photochemically controlled region below 200 km during the weak SC 24. While the intense space weather drivers, such as ICMEs, CIRs, Flares, and SEPs, have been shown to energize, deplete, and accelerate plasma in the topside ionosphere (e.g., B. M. Jakosky et al. 2015b; L. Ram et al. 2023; L. Liu et al. 2025), a unified picture of the topside ionospheric response over SC 24 and 25 is still missing. In particular, the evolution of the Martian topside ionosphere from the descending phase of SC 24 into the rising and maximum phase of SC 25 remains unexplored. To address this gap, we have tried to analyze the dayside and nightside observations from the northern hemisphere, where crustal magnetic fields are weak, throughout both cycles. Also, the extended coverage provided by MAVEN now offers, for the first time, a valuable opportunity to investigate the Mars topside ionosphere over different periods of solar activity during SC 24 and 25. It is to be noted here that the southern-hemisphere data are excluded to avoid strong crustal magnetic field effects.

This paper is structured as follows: Section 2 describes the data utilized in this work. Section 3 provides the detailed results on the Mars topside ionosphere. It is divided into two subsections: 3.1 and 3.2. Subsection 3.1 presents the observations on the topside ionosphere across different phases of SC over low- and mid-latitudes. Whereas, subsection 3.2 provides a focused comparative analysis of the topside ionosphere over different phases of SC 24-25. Section 4 discusses the observations, while Section 5 concludes the present study.

## 2. Data and Methods

To analyze the solar wind inputs and the topside plasma environment over Martian years (MY) 32-38 (2015 January 01 to 2024 November 14), multiple instruments aboard the MAVEN spacecraft (B. M. Jakosky et al. 2015a) have been utilized. The solar wind key parameters data, including the density, velocity, and dynamic pressure, are obtained from the MAVEN/Solar Wind Ion Analyzer (SWIA; J. S. Halekas et al. 2017). The SWIA measures the solar wind ions over a broad energy range of 5–25,000 eV and has a field of view of 360° × 90° with a resolution of ~22.5° (J. S. Halekas et al. 2017). The KP in-situ measurements of MAVEN/Magnetometer (MAG) are utilized to calculate the magnitude of interplanetary magnetic field (IMF; |**B**|) in MSO coordinates (J. E. P. Connerney et al. 2025). The Level 2, versions_19-22, revisions_01&02 (v19_r01; v20_r02; v21_r01; v22_r01) data of SWIA and MAG are being leveraged for the analysis (P. A. Dunn 2023). The electron energy flux data are obtained from the MAVEN/Solar Wind Electron Analyzer (SWEA; D. Mitchell et al. 2016). The SWEA is an electrostatic analyzer designed to measure solar wind electrons and ionospheric photoelectrons in the Martian environment, covering an energy range of 3 to 4,600 eV. The SWEA Level 2, versions_19-22, revisions_01&02 (v19_r01; v20_r02; v21_r01; v22_r01) data are leveraged for the analysis (D. Mitchell et al. 2016; P. A. Dunn 2023).

The solar irradiance, including the soft X-ray (SXR) and Extreme Ultraviolet (EUV) flux, is obtained using the MAVEN/Extreme Ultraviolet

Monitor (EUVM; F. G. Eparvier et al. 2024), which is a part of the Langmuir Probe and Waves (LPW; L. Andersson et al. 2017) onboard the MAVEN spacecraft. The MAVEN/EUVM Level 3, daily averaged modelled irradiance spectra (F. G. Eparvier 2024; E. M. Theimann et al. 2017) are used to produce time series variation of solar irradiance including SXR and EUV fluxes from 2015 to 2024. The SWIA and EUVM data are accessed through the NASA Planetary Data System (PDS) and by utilizing the Python Data Analysis and Visualization tool (PyDIVIDE; MAVEN SDC et al. 2020). Along with the solar irradiance flux, the monthly averaged Sunspot Numbers (SSNs) data are used to categorize the different phases of SC 24 and 25. The SSNs data are obtained from the WDC-SILSO, Royal Observatory of Belgium, Brussels (https://doi.org/10.24414/stcz-kc45).

The MAVEN spacecraft usually takes measurements of the Mars topside ionosphere during inbound-periapsis-outbound phases while traversing between 150 and 500 km altitudes above the Mars surface. We have utilized the MAVEN/Neutral Gas and Ion Mass Spectrometer (NGIMS) instrument data to obtain the ions ($O^+$, $O_2^+$, $CO_2^+$, $NO^+$, $C^+$, $N^+$, & $OH^+$) density (P. R. Mahaffy et al. 2015). The MAVEN/NGIMS measures the ions and neutrals of the Martian atmosphere/ionosphere using a dual ion source and a quadrupole mass analyzer. The Level 2, version_v08, revision number_r01 (v08_r01) data of NGIMS (M. K. Elrod, M. Benna & T. Navas 2014) are utilized in our analyses. The Langmuir Probe and Waves (LPW) instrument data are used to obtain the electron density and temperature measurements (L. Andersson 2017). The LPW Level 2, versions_19-22, revisions_01&02 (v19_r01; v20_r02; v21_r01; v22_r01) data (L. Andersson 2017; P. A. Dunn 2023) are used in the present work. The MAVEN datasets were accessed through the NASA Planetary Data System (PDS), MAVEN Science Data Center (SDC), and using the Python Data Analysis and Visualization tool (PyDIVIDE). In addition, the present study have utilized the NASA's Community Coordinated Modeling Center (CCMC) Space Weather Database (https://kauai.ccmc.gsfc.nasa.gov/DONKI/search/), MAVEN Science Data Centre (SDC; https://lasp.colorado.edu/maven/sdc/public/) and various catalogs (H. Huang et al. 2019; D. Zhao et al. 2021; P. Geyer et al. 2021; L. Ram et al. 2023) to remove the intense space weather periods (ICMEs, CIRs/SIRs, Flares, SEPs), crustal fields effects (J. E. P. Connerney et al. 2005), and dust storm periods (S. D. Guzewich et al. 2020; C. Martín-Rubio et al. 2024) from the MAVEN in-situ data.

## 3. Results

In this study, a comprehensive analyses of the Mars topside ionosphere (150-500 km altitude) are conducted between MY 32 and 38 (2015-2024) over the northern hemisphere (0-60°N). The analyses are based on MAVEN in-situ observations spanning from the declining and minimum phases of SC 24 (2015-2019) to the ascending and maximum phases of SC 25 (2020-2024). We first removed all major energetic solar transients' event periods such as ICMEs, CIRs, and Solar Flares (NASA CCMC; MAVEN SDC, as mentioned in Section 2) from the MAVEN data in order to avoid their effects on the Mars topside ionosphere. Additionally, the data used in this study exclude the periods of dust storms (S. D. Guzewich et al. 2020; C. Martín-Rubio et al. 2024) as well as free from the influence of strong crustal magnetic fields (J. E. P. Connerney et al. 2005). The diverse suite of observations from the MAVEN instruments are categorized into four phases of SC, then, further divided into low- (0-30°N) and mid-latitude (31-60°N) datasets over the dayside (SZA < 60°) and nightside (SZA > 115°). The MAVEN NGIMS and LPW inbound in-situ data are being utilized to conduct this work.

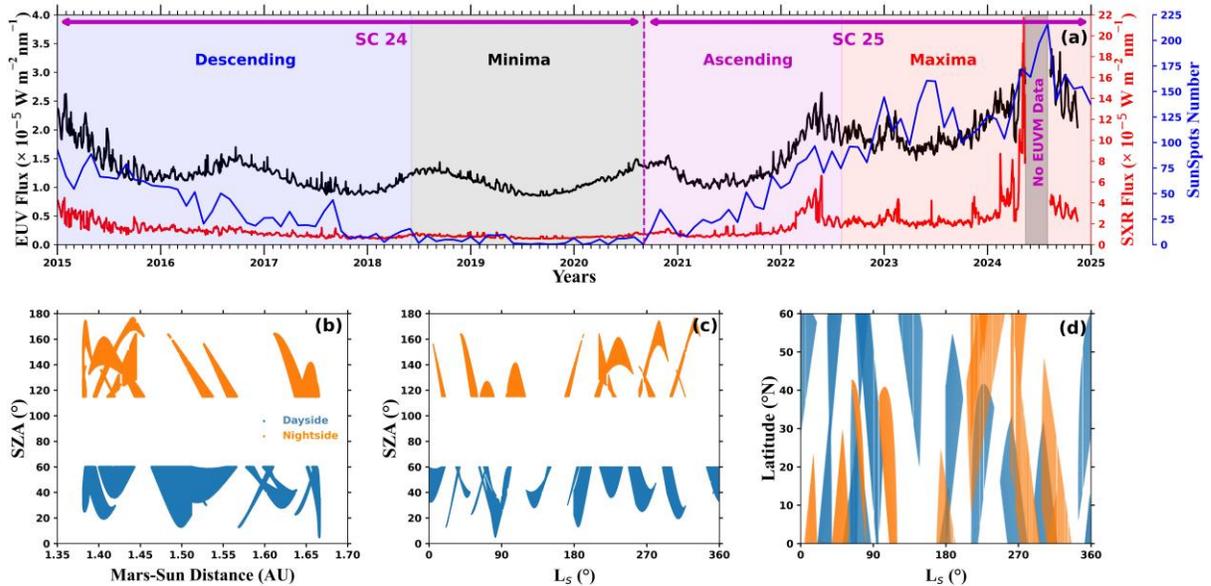

**Figure 1.** (a) Top Panel shows a time series of extreme ultraviolet (EUV) flux (left y-axis, black color), soft x-ray (SXR) flux (right y-axis, red color), and sunspots numbers (SSNs) (right y-axis, blue color)-reveals solar activity trends with Solar Cycles 24 and 25. The color shaded regions depicted in different color represent the different phases of Solar Cycles. The vertical dotted magenta line indicates the start of SC 25. The grey shaded area during the year 2024 indicates the non-availability of MAVEN/EUVM data. (b-d) Bottom panel shows the MAVEN periapsis observations from Martian years (MY) 32 to 38 (2015-2024) over different solar zenith angles (SZAs), Mars-Sun distances (au), Martian latitudes, and Solar longitudes (Ls) on both the dayside (blue color) and nightside (orange color).

The top panel (Figures 1a) illustrates the time-series variation of integrated EUV flux (11-120 nm; shown in black; left y-axis), SXR flux (1-10 nm; shown in red; right y-axis), and the variation of monthly averaged sunspots numbers (SSNs; shown in blue, right y-axis) during SC 24 and 25 (2015-2024). The two-headed arrows in the magenta color represent the MAVEN observational periods during the SC 24 and SC 25. The EUV flux magnitude (black; Figure 1g) decreases from approximately $2.5\times10^{-5}$ W m$^{-2}$ nm$^{-1}$ at the beginning of 2015 (descending phase SC 24) to around $1\times10^{-5}$ W m$^{-2}$ nm$^{-1}$ by the end of the year 2019 (solar minima of SC 24). Following this, the solar EUV flux increases, reaching its peak magnitude of $3.8\times10^{-5}$ W m$^{-2}$ nm$^{-1}$ in June 2024 (solar maxima of SC 25). In addition, the SC variability is evident in the time-series of SSNs. The monthly SSN decreases from 93 (Jan, 2015) to nearly 1 by the end of 2019, up to 2020, indicating the minima phase of SC. Following this, the SSNs started increasing and approaching a maximum value of nearly 225, indicating the solar maxima period. The MAVEN observations are categorized into four different solar active periods of SC 24 and 25 (Figure 1a) according to SSNs criteria (F. Ouattara & C. Amory-Mazaudier 2009; S. Sawadogo et al. 2023).

The different solar active periods of SC 24 and SC 25 (2015-2024) are categorized as follows:

1. **SC 24 Descending Phase:** 2015 January 01 to 2018 June 01 [SSNs: 16-93]
2. **SC 24 Minima Phase:** 2018 July 01 to 2020 September 01 [SSNs: 1.6-0.6]
3. **SC 25 Ascending Phase:** 2020 October 01 to 2022 August 01 [SSNs: 14-75]
4. **SC 25 Maxima Phase:** 2022 September 01 to 2024 November 14 [SSNs > 96]

In addition, Figures 1b-d represent the observational span of the MAVEN monitoring of the topside ionosphere at different solar zenith angles (SZA) and northern latitudes with respect to different Mars-Sun distance (au) and solar longitudes (Ls) in both the dayside (shown in blue) and nightside (shown in orange).

### 3.1 Mars topside ionospheric behavior during different phases of SC 24 and 25

Figure 2 illustrates the topside ionospheric behavior over low- and mid-latitude regions during the descending phase of SC 24. The top and bottom panels show the dayside and nightside observations, respectively. The first two columns show the variations in electron temperature and density, while the subsequent columns present the ion species profiles ($CO_2^+$, $O_2^+$, $NO^+$, $OH^+$, $O^+$, $N^+$ & $C^+$) arranged in order of their decreasing mass (this format is consistently applied across Figures 2-6). The low- and mid-latitude median profiles are depicted in black and magenta color, respectively. The electron and ions profiles were generated by taking the median over a 10 km altitudinal bin of density (temperature) along with the median absolute deviation (MAD; shown by the colored horizontal bars). On the dayside, both the electron and ions densities are higher at low-latitude, particularly above 250 km altitude, compared to mid-latitude (Figures 1b & Figures 1d-i), while the electron temperature and $CO_2^+$ density show no significant variation (Figures 1a&c). In addition, the peak height and density of the lighter ions ($OH^+$, $O^+$, $N^+$ & $C^+$) remain nearly similar in both low- and mid-latitudes. On the nightside, the median electron temperature is higher at all altitudes over mid-latitude compared to low-latitude (Figure 2j), while the electron density decreases with altitudes in mid-latitude, particularly above 200 km (magenta; Figure 2k). The previous findings (R. E. Ergun et al. 2015; M. F. Vogt et al. 2017; F. González-Galindo et al. 2021) suggested that the latitudinal variation in electron temperature and the background neutral atmosphere can affect the electron density at and above the peak. The increased electron temperature can be expected due to decreased neutral densities above 200 km that can cause less frequent collisions (C. M. Fowler et al. 2015). Figures 2k-r show that the ions densities in the mid-latitude region are higher in magnitude below 200 km altitude. However, due to the unavailability of ion data in the low latitude region between 200 and 350 km, an inference cannot be drawn. Although, the enhanced nightside electron temperature especially over mid-latitude can associate with a lesser electron density (Figures 2j-k), which is consistent with the previous findings as mentioned above.

Figure 3 shows the topside ionospheric measurements during the minima phase of SC 24. The electron and ion species show higher densities at the low-latitude region than at the mid-latitude on both dayside and nightside. On the dayside, the difference is prominent above 300 km altitude for heavier ions ($CO_2^+$, $O_2^+$, $NO^+$), whereas, for lighter ions, it occurs above and below the peak. The peak appears to be unaltered for both dayside and nightside.

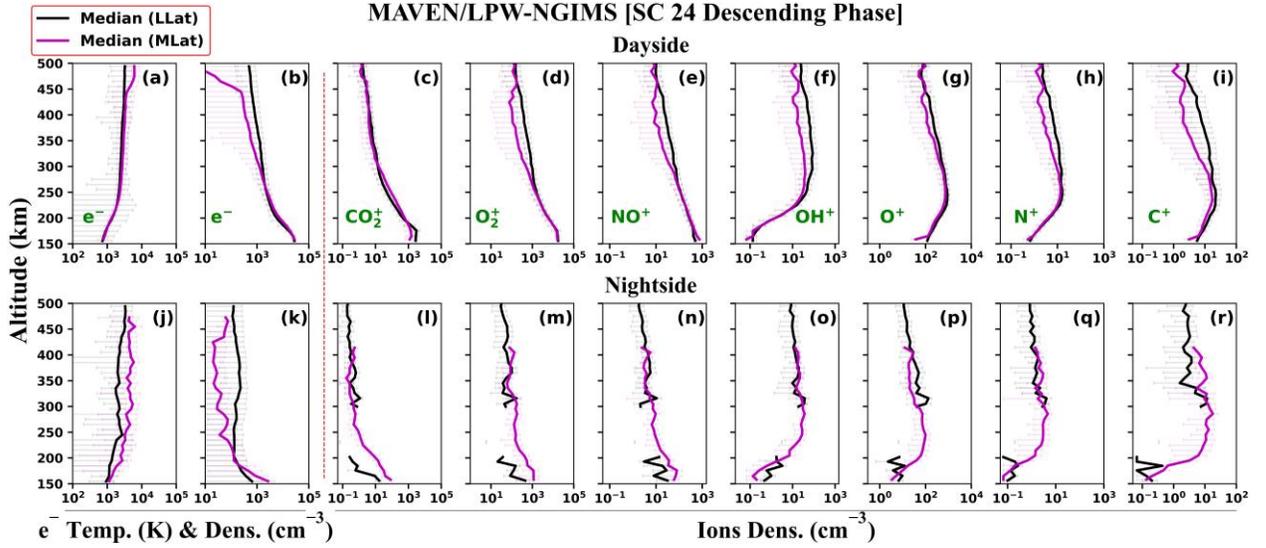

**Figure 2.** The LPW and NGIMS observations of the topside ionospheric median electron density and temperature, and ion density ($CO_2^+$, $O_2^+$, $NO^+$, $OH^+$, $O^+$, $N^+$ & $C^+$) as a function of altitude in both dayside (top panel; **a-i**) and nightside (bottom panel; **j-r**) at low- (0-30°N, black profile) and mid-latitude (31-60°N, magenta profile) regions during descending phase of solar cycle (SC) 24. The notations such as Temp. and Dens. are referring to the temperature and density, respectively. The red dotted vertical lines divide the electron and ions measurements. The abbreviations: LLat and MLat mentioned in legend box (top left corner, red color box) are referring to the low-latitude and mid-latitude, respectively.

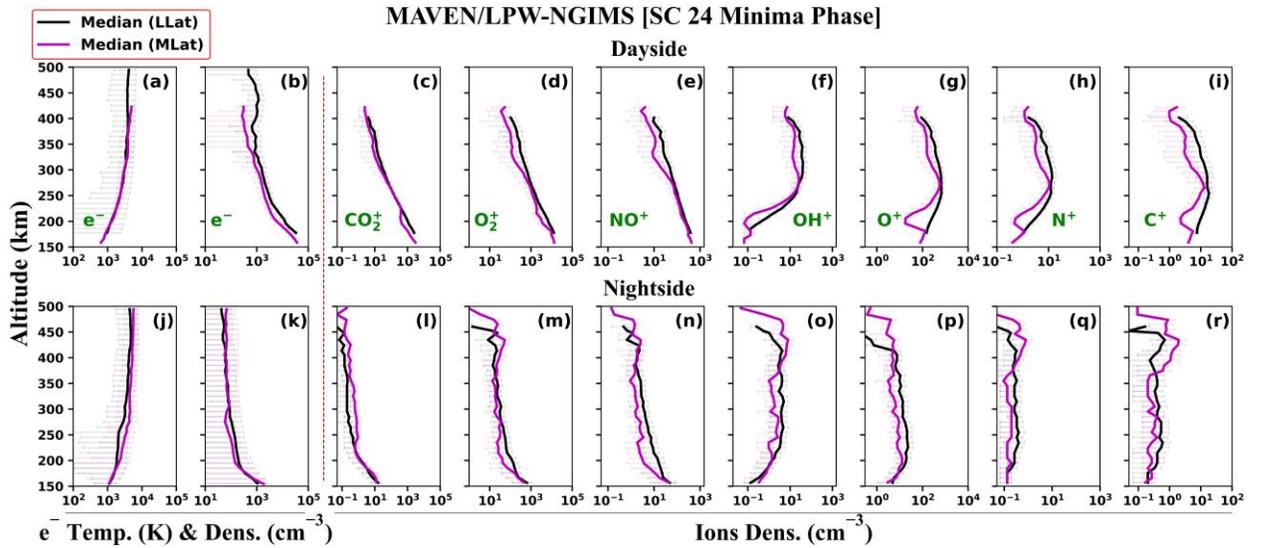

**Figure 3.** Observations of the Mars topside ionosphere during the minima phase of SC 24 (remaining captions are similar to Figure 2).

Figure 4 illustrates the variation of topside ionosphere during the ascending phase of SC 25. During the ascending phase of SC 25, the dayside electron and ion species show slightly higher densities in the low-latitude region than mid-latitude below 300 km altitude. On the nightside, the $O^+$ shows higher density in mid-latitude, although a significant difference has not been observed for other species. It is important to note that due to increase in the MAVEN's periapsis altitude after August 2020 (from ~150 to ~180-220 km; M. Elrod & M. Benna 2022), variations below 200 km cannot be examined during this period.

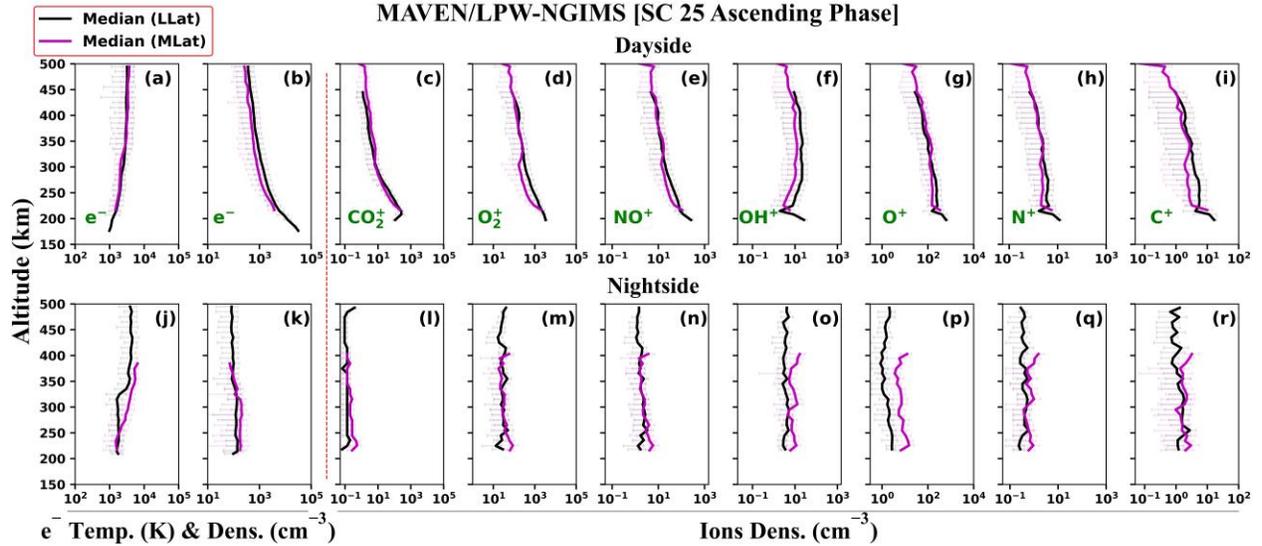

**Figure 4**. Observations of the Mars topside ionosphere during ascending phase of SC 25 (remaining caption is similar to Figure 2).

Figure 5 shows the topside ionospheric variation during the maxima phase of SC 25. During the ongoing maxima phase of SC 25, the dayside electron and heavier ions ($O_2^+$, $CO_2^+$) species exhibit a notable density difference below 400 km altitude, with significantly higher magnitude observed in the low-latitude region compared to the mid-latitude region. Whereas, for lighter ions ($OH^+$, $O^+$, $N^+$, & $C^+$), the densities are significantly higher below 300 km altitude in the mid-latitude compared to the low-latitude region. Although, the peak altitudes of lighter ions occur at lower altitude over mid-latitude than low-latitude. For example, the $O^+$ ion has a peak height of 330 km in the low-latitude, whereas, it exists at nearly 250 km in the mid-latitude region. In addition, no significant change is observed in the nightside (due to the less data coverage over mid-latitude).

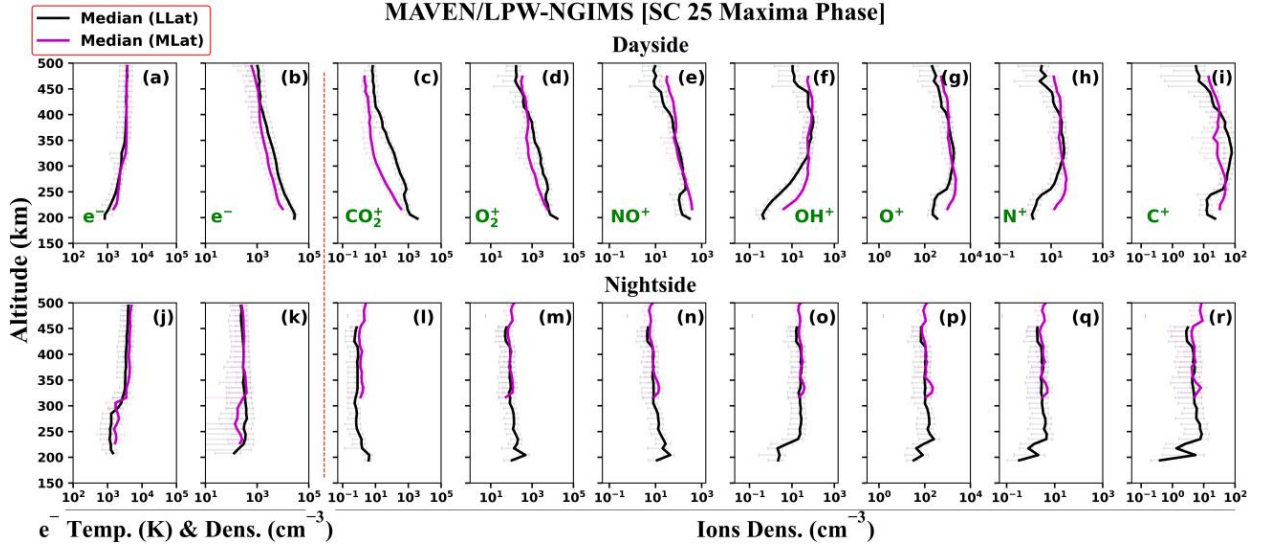

**Figure 5**. Observations of the Mars topside ionosphere during the maxima phase of SC 25 (remaining caption is similar to Figure 2).

The observations shown in Figures 2-5 present a significant difference in the density profiles of the Mars topside ionosphere over low- and mid-latitude regions across different phases of SC 24 and SC 25. The previous work (Z. Girazian et al. 2019; P. Withers et al. 2023; K. Hensley & P. Withers 2021; K. Hensley et al. 2022; Z. Girazian et al. 2023; V. Mukundan et al. 2025) have reported a significant impact of solar radiation and particle events on the ionosphere during the declining phase of SC 24. The work by B. Sánchez-Cano et al. (2016) has suggested a 7% and 8% lower peak density and peak altitude, respectively during low active phase compared to high solar activity during SC 23. Further, the study by K. Hensley & P. Withers (2021), which utilized the radio occultation measurements along with a photochemical equilibrium model, identified a strong correlation between the increased solar irradiance and enhanced electron densities throughout the photochemical region (~100-180 km), specifically both above and below the peak (at SZAs ~76-78°). They suggested a strong role of solar irradiance in altering the ionospheric densities rather than electron temperature and minor ions such as $NO^+$ and $N_2^+$ below 180 km altitudes. However, the response of the topside ionosphere to varying solar cycle phases remain poorly understood. Furthermore, the absence of the data across multiple SC phases has prevented a detailed understanding of the Martian ambient space environment, particularly in the region above the photochemical region (100-180 km). In the following analysis, we investigate this aspect, by comparing the behavior of Mars topside ionosphere across different phases of SC.

## 3.2 Comparison of the Martian Northern Plasma density profiles across four phases of Solar Cycles 24 and 25

The behavior of the Martian northern topside ionosphere is examined across different phases of SC 24 and SC 25, highlighting the variations in the electron and ions species ($CO_2^+$, $O_2^+$, $NO^+$, $OH^+$, $O^+$, $N^+$ & $C^+$). Figure 6 illustrates a comparison of the median plasma density profiles on both dayside (top panel; Figures 6a-i) and nightside (bottom panel; Figures 6j-r). The descending and minima phases of SC 24, and the ascending and maxima phases of SC 25 are depicted by the blue & black, and magenta & red color solid profiles, respectively. A significant difference can be noted in the electron and ions species across four phases of SC. It is evident from Figure 6 that the plasma density approaches its peak magnitude during the maxima phase of SC 25 (red) compared to the remaining phases in both the dayside and nightside. The magnitude of plasma density is 1 to 2 order higher on the dayside compared to the nightside during each SC phases.

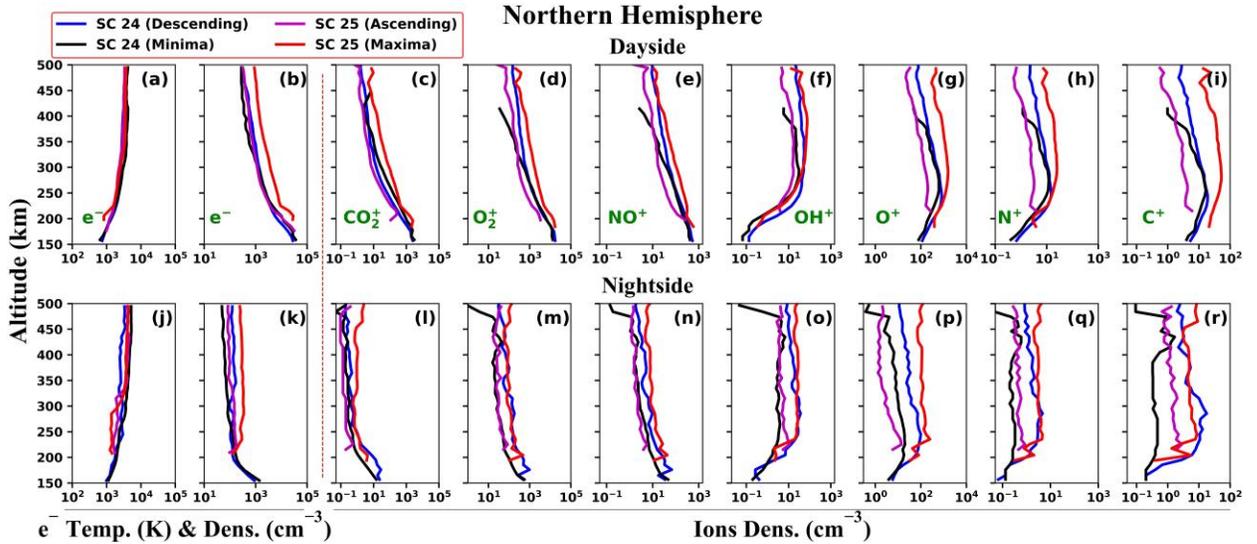

**Figure 6**. The northern hemispheric (0-60°N) median plasma density ($e^-$, $CO_2^+$, $O_2^+$, $NO^+$, $OH^+$, $O^+$, $N^+$ & $C^+$) profiles in both dayside (top panel; a-i) and nightside (bottom panel; j-r) during the different phases of SC 24 and 25. The abbreviation SC stands for Solar Cycle. The descending, minima, ascending, and maxima phases of SC are represented by solid lines in blue, black, magenta, and red color, respectively.

On the dayside, the electron temperature (Figure 6a) across different phases remains similar above 300 km. Although, during maxima phase, it is lesser below 300 km altitude compared to other phases. The electron density magnitude remains nearly similar up to 300 km across the descending, minima, and ascending phases. However, above 300 km, it increases from the minima phase through the descending and ascending phases, reaching its maximum magnitude during the maxima phase. This indicates that at higher altitudes, the variation in electron density can directly correlates with the varying EUV across descending to maxima phases of SC 24 and 25. During the maxima phase (red profile; Figure 6b), the electron density is consistently higher at all altitudes compared to other phases. The electron density (Figure 6b) enhanced by 2 to 5 times compared to other phases. Additionally, for ions, the lowest plasma density magnitude is observed during ascending phase (magenta; Figures 6c-i) and increasing across minima, descending, and approaches to peak magnitude during maxima. Whereas, the ions densities during minima and descending phases are nearly similar, especially below 300 km altitude, except for the $CO_2^+$ ion. This indicates a slightly different response of the ions, with lowest density occurring during ascending phase, unlike electron, which exhibit their lowest density during the minima phase. This is likely primarily driven by the combined effects of EUV flux, ionospheric chemistry, and solar wind (J. L. Fox 2009; T. E. Cravens et al. 2017). During maxima phase, the $CO_2^+$ and $O_2^+$ ions (Figures 6 c&d), show enhancements ranging from 1.2 to 13 times and 1.1 to 12 times, respectively (indicated in red), when compared to low solar activity phases (represented in blue, magenta, and black). For $O^+$ ion, the density increased by 1.1 to 14 times, whereas, the peak density increased by 1.9 to 2.5 times compared to the remaining phases. During the maxima phase, the peak of the lighter ions ($OH^+$, $O^+$, $N^+$ & $C^+$) occurs at higher altitude compared to other phases. For example, the $O^+$ peak is uplifted by nearly 40-50 km altitude (Figure 6g) compared to the low active phases.

   On the nightside, a similar trend of increased electron and ions densities is observed during the maxima phase compared to other phases. The electron and ions densities approach its peak magnitude during maxima phase, followed by progressively lower magnitude during the ascending, descending, and minima phases. The electron density at various altitudes increases by approximately 1.1 to 5 times during maxima phase compared to other phases of SC (Figure 6k). However, the difference in $O_2^+$ densities (represented by the blue and red curves) between the descending and maxima phases is not particularly significant. Nevertheless, a noticeable increase, ranging from 1 to 4.5 times is observed when compared to minima and descending phases. Additionally, a pronounced difference is observed for the $O^+$ ion, where the density difference increases with altitude. The peak density increases by nearly 2.5 to 12 times compared to low active phases. Unlike the dayside, the peak altitude of lighter ions ($OH^+$, $O^+$, $N^+$ & $C^+$) remain relatively consistent across all phases.

## 4. Discussion

The climatology of Martian topside ionosphere is presented across different phases of Solar Cycles 24 and 25. The observations shown in Figures 2-5 present a significant difference in the density profiles of the Mars topside ionosphere over low- and mid-latitude regions across different phases of SC 24 and SC 25. The elevated plasma densities in the low-latitude on the dayside result from their proximity to the subsolar point, where increased EUV and SXR flux heat and expand the neutral atmosphere to higher altitudes, leading to enhanced ionization in the topside ionosphere (B. Sánchez-Cano et al. 2015; J. L. Fox, P. Zhou & S. W. Bougher 1996). Figures 6 exhibits enhanced dayside ionospheric densities during maxima phase of SC 25, primarily due to the enhanced solar radiation, specifically the EUV and SXR fluxes, which are responsible for the photoionization of neutrals (primarily $CO_2$; a major neutral species on Mars) at these altitudes (M. Benna et al. 2015; P. R. Mahaffy et al. 2015; T. E. Cravens et al. 2017). The EUV and SXR fluxes during SC 25 maxima phase is ranging from 2.1–3.5 × $10^{-5}$ W $m^{-2}$ $nm^{-1}$ and 2.5-22 × $10^{-5}$ W $m^{-2}$ $nm^{-1}$, respectively, which is nearly 1.4-2 times (EUV) and 1.5-6 times (SXR) compared to the low solar active phases. The increasing radiation flux causes atmospheric heating, which leads to an inflated neutral atmosphere, resulting in an increased neutral density (M. Pilinski et al. 2019; K. Hensley & P. Withers 2021). The elevated neutral density (primarily composed of $CO_2$), results in enhanced photo-ionization (M. Pilinski et al. 2019) along with the secondary ionization driven via the electron impact (which contributes ~10% to total ionization; T. E. Cravens et al. 2017) process in the topside ionosphere. This increases the ionization rate and leads to elevated plasma density on the dayside. The previous work by K. Hensley, P. Withers & E. M. Theimann (2022) have demonstrated a similar enhancement at a given altitude of the dayside ionosphere below 25° SZA during a high solar irradiance period (17–25 April 2015) compared to low solar irradiance period (20–23 October 2017). However, the present study, includes an extended observational span from 2015 to 2024, incorporating the different solar activity periods, which will enhance our understanding about Martian topside ionosphere during SC 24 and 25. Additionally, the elevated peak of $O^+$ ion (SC maxima) primarily depends upon the $O/CO_2$ ratio (Z. Girazian et al. 2019). The previous work (N. Yoshida et al. 2021; K. Hensley, P. Withers & E. M. Theimann 2022) applied on small datasets indicated that the increment in $O/CO_2$ ratio during periods of higher solar activity leads to an elevated O$^+$ peak, which are consistent with the present study results accompanied large number of datasets over different solar activity periods.

We found that the dayside and nightside electron temperature is minimal during SC maxima phase below 300 km altitude (Figures 6a&j) compared to lower active phases. This can be due to the increasing electron cooling rate via electrons collisions with the inflated neutral atmosphere (M. Pilinski et al. 2019) during higher solar irradiance periods. Although, above the 300 km, the increasing electron temperature with respect to the altitude during the SC maxima phase is showing a similar magnitude as observed for the lower active phases. As the neutral density decreases exponentially above 180 km altitude, which results in infrequent or lesser collisions of electron with the neutrals (C. M. Fowler et al. 2015). Moreover, the neutral atmosphere within which the ionosphere is embedded, energized by the increased EUV and SXR flux, plays a significant role in altering the vertical topside ionospheric structure (R. E. Ergun et al. 2015; M. F. Vogt et al. 2017; F. González-Galindo et al. 2021).

Besides the direct influence of solar irradiance, the Martian topside ionosphere is also modulated by other factors such as solar wind dynamic pressure (SWDP), interplanetary magnetic field (IMF), and solar energetic electron flux. The energetic solar wind particles can energize, increase ionization, and eventually lead to plasma escape from the topside ionosphere (B. M. Jakosky et al. 2015b; L. Ram et al. 2023; Y. Chi et al. 2023; L. Liu et al. 2025) into space. Albeit, in this study, we have removed all the observations during ICMEs, CIRs, and Flares to provide a solar quiet-time scenario during the different phases of solar cycles. To provide an overall time-series dependence of top ionosphere on the solar particle, field, and radiation, we have studied the variation of the total averaged electron and ions density above 300 km altitude across varying phases of SC 24 and 25. Figure 7 illustrates the time-series variations of averaged solar EUV (top panel) and upstream solar wind parameters [Second to Fourth panels: SWDP (nPa), resultant IMF (|**B**| (nT)), and electron energy flux (100-500 eV); shown in left y-axis] across different phases of SC 24 and 25. The variation of monthly averaged dayside electron (right-side; green colored y-axis) and ion (right-side; purple colored y-axis) densities are illustrated over different phases of SC 24 and 25. It is evident that the averaged electron and ion densities closely follow the variations in EUV, IMF (|**B**|), and solar energetic electron flux across different phases of SC. Particularly, the averaged electron density decreases from the descending phase to its lowest levels during the minima phase, then increases through the ascending phase, reaching a maximum during maxima phase (also shown in vertical profile of Figures 6b&k).

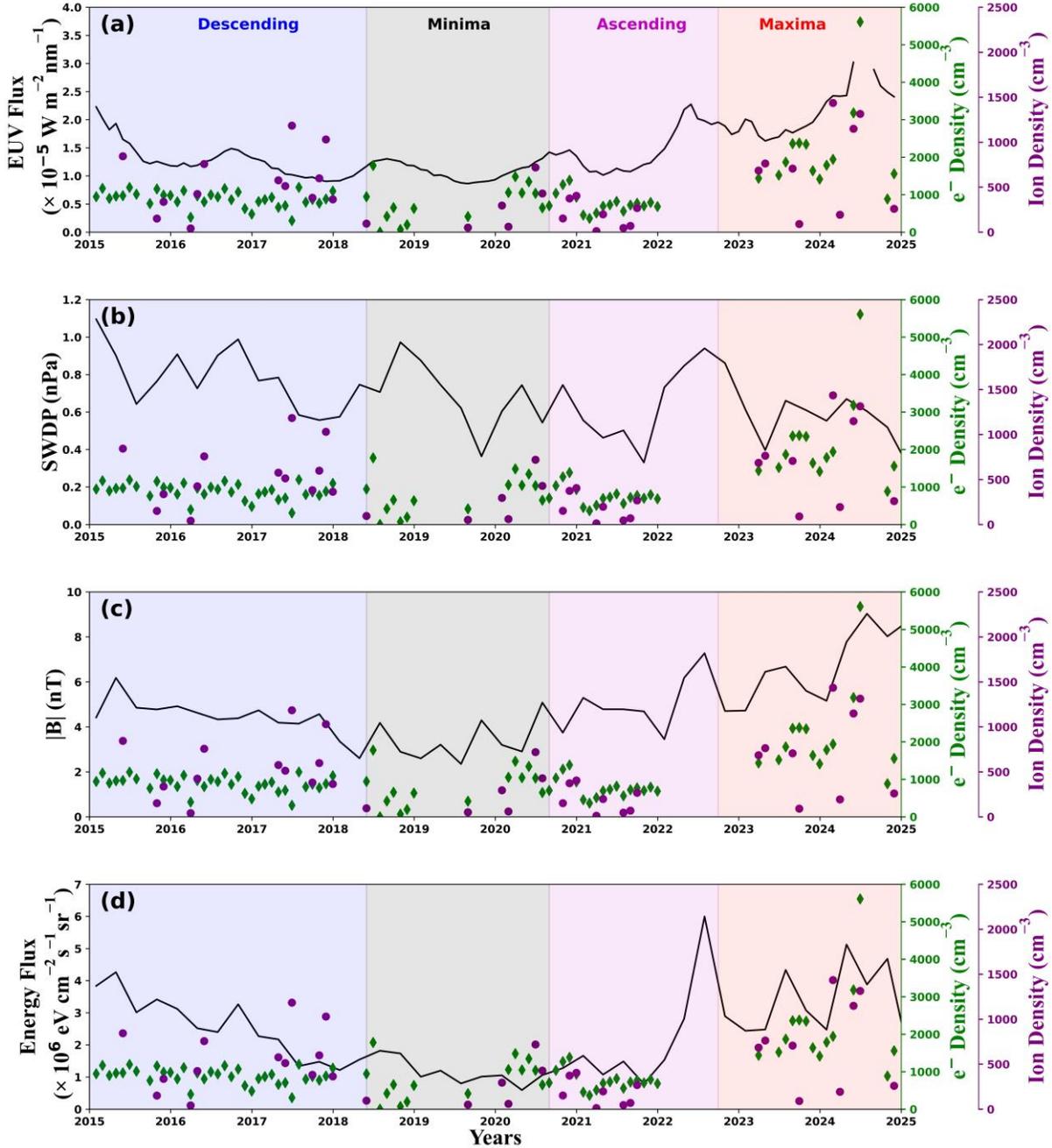

**Figure 7.** Time series variations of averaged solar EUV flux (Top panel; left y-axis) and upstream solar wind dynamic pressure (SWDP (nPa)), resultant IMF (|**B**| (nT)), and electron energy flux (100-500 eV) (Second to Fourth Panels, left y-axis) over different phases of Solar Cycles 24 and 25 in 2015-2024. Similarly, ionospheric monthly averaged electron (green diamonds) and ion density (purple dots) variations (>300 km altitude) are illustrated in the first and second right y-axis, depicted in green and purple color font styles, respectively.

This suggests that the topside averaged electron density exhibits a strong dependence on the EUV flux along with a noticeable variation with the IMF magnitude (|**B**|) and solar electron flux; while showing minimal correlation with SWDP over the longer time-scale of the SC. In addition, the ion density also shows a similar trend across the progressing phases of SC. However, the ion densities (purple dots) appear more scattered compared to electron densities (green diamonds) values, suggesting the influence of ionospheric chemistry or transport processes in shaping the ions distribution. The study by Sánchez-Cano et al. (2015) demonstrated that during the lower solar active phases, the solar wind induced magnetic field can penetrate deeper to the lower altitudes, control the ionosphere and reduce the vertical extent of photochemical region. Our results align with previous findings that show a less dense topside ionosphere during periods of low solar activity. In contrast, higher densities during the maximum phase indicate a more extended ionosphere, primarily due to increased heating from elevated EUV and SXR fluxes as explained in the preceding paragraph. The extended ionosphere, therefore, reduces the deeper penetration of the induced magnetic field lines on the dayside. This demonstrates that solar irradiance flux is a dominant factor in regulating the topside ionosphere over long time-scales, such as solar cycles, along with the significant effect induced by the IMF and solar energetic electron flux.

On the other hand, the nightside ionosphere, which is primarily seeded or controlled via the day-to-night plasma transport (P. Withers et al. 2012; T. E. Cravens et al. 2017; Z. Girazian et al. 2019), as well as the electron impact driven by the solar wind energetic particles or dayside photoelectrons (R. J. Lillis et al. 2009; R. J. Lillis & D. A. Brain 2013; S. Xu et al. 2016, 2017). The higher dayside density act as a reservoir that supplies more ions through plasma transport from day-to-night due to the pressure gradient and can cause higher density in the top nightside ionosphere during maximum solar activity compared to the low/moderate solar activity periods (depicted in blue, magenta, and black; Figure 6). The transport mechanism seems to work more efficiently for the lighter species compared to molecular species such as $CO_2$, which could cause a minimal difference (Figures 6i-j) in the densities of $O_2^+$ and $CO_2^+$ ions during different periods of SC (S. W. Bougher et al. 2015; M. K. Elrod et al. 2017). However, the previous studies have reported that the transport of plasma from day-to-night is possible up to 110° SZA, whereas, the nightside data considered in this work lie above 115° SZA. To decipher and address this aspect, we have analyzed the second important factor i.e., the solar wind particle precipitation or photoelectron-driven ionization, which can influence the nightside ionospheric structure. We calculated the electron flux (100-500 eV and 500-1000 eV) using the MAVEN/SWEA instrument data. The averaged electron flux is nearly $5-7\times10^7$ eV cm$^{-2}$ s$^{-1}$ sr$^{-1}$ and $4\times10^5$ eV cm$^{-2}$ s$^{-1}$ sr$^{-1}$, respectively during SC maxima phase and show a percentage deviation of nearly 66% and 33%, respectively from the low solar active periods. Furthermore, the higher induced magnetic field during maxima phase (Figure 7c) results in more draping of the filed lines around ionosphere, that gives a pathway to energetic electrons to increase ionization via the electron impact on the nightside neutral species. This results into an elevated plasma density in the topside nightside ionosphere compared to other low active periods of the SC 24 and 25.

The above discussion presents an overview of the Martian northern hemispheric topside ionosphere, where the effects of the crustal field are minimal or absent. However, the southern hemisphere, presents an interesting scenario to study the topside ionosphere, in the presence of strong crustal fields during varying phases of solar cycles (SC 24 and 25). In addition, the changing seasons on Mars across MY 32-38 will also present an open facet to understand its topside ionospheric variability with respect to the seasons. This work is currently underway and will be communicated in the near future.

## 5. Conclusions

The Martian northern topside ionosphere is investigated across different phases of Solar Cycles 24 and 25 using MAVEN mission datasets. The northern plasma density profiles have been analysed and their behavior is studied with respect to the solar irradiance and solar wind parameters. The following points concluded this study:

1. The Martian topside ionosphere exhibits significant variability across Solar Cycles 24 and 25, with stronger variation over low-latitude due to more direct solar radiation compared to mid-latitudes.
2. The plasma density profiles exhibit a pronounced variations and densities are 1-2 order higher in magnitude on the dayside than nightside across the different phases of solar cycles 24 and 25. The largest plasma densities have been observed during the maxima phase (red color; Figure 6) compared to low-active periods of Solar Cycles.
3. Electron density:

a. Decreases from descending to minimum phase.
   b. Increases during the ascending phase, approaches to its maximum magnitude above 300 km during solar maxima phase.
   c. Increases by factors of 2-5 during maximum phase compared to low active phases.
4. Ion density:
   a. Decreases from descending to minimum phase, reaching lowest magnitude during the ascending phase and approaches to maximum magnitude during solar maxima period.
   b. Increases by factors of 1.2–14 during solar maxima compared to other low active phases.
   c. For lighter ions, the peak is elevated on the dayside during the maxima phase compared to low active phases, whereas, on the nightside, the peak heights remain stable across different phases.
   d. Particularly, the lighter ion such as $O^+$, which is one of a major ion of the topside ionosphere, exhibits an elevated peak density of 1.9–2.5 times, with a 40–50 km rise in peak altitude during solar maxima compared to other phases.

5. The increased dayside plasma densities are attributed to the increased EUV flux (approximately 1.4-2 times higher) and SXR flux (1.5-6 times higher) along with the IMF and solar wind electron flux relative to other low-activity periods of Solar Cycles.

6. On contrary, the enhanced nightside densities are likely driven by elevated integrated electron flux and day-to-night plasma, induce by a strong pressure gradient between the dayside and nightside.

In summary, this study utilizing the diverse suite of MAVEN data over 10 Earth years (nearly 6.5 Martian years), provides an overall picture of the Solar Cycle-dependent topside ionospheric variation and will serve as a valuable quiet time reference data for scientific community in investigating the space weather-induced perturbations.

## Acknowledgments

L. Ram, C. Singh, and A. R. Sharma acknowledge the fellowship from the Ministry of Education, Government of India for carrying out this research work. We sincerely acknowledge the NASA PDS and the MAVEN team, especially the LPW, NGIMS, SWIA, SWEA, and EUVM, for the data sets. All the datasets from MAVEN are available at the PDS Planetary Plasma Interactions Node

(https://pds-ppi.igpp.ucla.edu/mission/MAVEN) and MAVEN SDC (https://lasp.colorado.edu/maven/sdc/public/), specifically, the LPW data (L. Andersson 2017; P. A. Dunn 2023), the NGIMS data (M. K. Elrod et al. 2017; M. Benna & T. Navas 2014), the SWIA data (J. S. Halekas 2017; P. A. Dunn 2023), the EUVM data (F. G. Eparvier et al. 2024), and the SWEA data (D. Mitchell et al. 2016). The SWIA, MAG, and SWEA data are accessed using the Python Data Analysis and Visualization tool (PyDIVIDE; MAVEN SDC et al. 2020). The monthly SSNs data are obtained from the WDC-SILSO, Royal Observatory of Belgium, Brussels (https://doi.org/10.24414/stcz-kc45). We also acknowledge the NASA's CCMC Space Weather Database of Notifications, Knowledge, and information (DONKI; https://kauai.ccmc.gsfc.nasa.gov/DONKI/search/) for the identification of space weather events. This work is also supported by the Ministry of Education and Department of Space, Government of India.

**Supporting Data**

The derived data products utilized and produced during this work can be found in the Zenodo repository (L. Ram et al. 2025).**References**

Andersson, L. 2017, urn:nasa:pds:maven.lpw::1.2, PDSS, doi: 10.17189/1410658

Benna, M., Mahaffy, P. R., Grebowsky, J. M., et al. 2015, GeoRL, 42, 8958

Bougher, S. W., et al. 2017, in The Atmosphere and Climate of Mars, ed. R. M. Haberle, R. T. Clancy, F. Forget, M. D. Smith, & R. W. Zurek (Cambridge: Cambridge Univ. Press), 433

Bougher, S. W., Pawlowski, D., Bell, J. M., et al. 2015, JGRE, *120*, 311

Chaufray, J. Y., Gonzalez-Galindo, F., Forget, F., et al. 2014, JGRE, 119, 1614

Chi, Y., Shen, C., Cheng, L., et al. 2023, ApJS, 267, 3

Cloutier, P. A., McElroy, M. B., & Michel, F. C. 1969, JGR, 74, 6215

Connerney, J. E. P., Acuña, M. H., Ness, N. F., et al. 2005, PNAS, 102, 14970

Connerney, J. E. P. 2025, urn:nasa:pds:maven.mag.calibrated::2.34, PDSS, doi: 10.17189/1414178

Cravens, T. E., Hamil, O., Houston, S., et al. 2017, JGRA, 122, 626

Dunn, P. A. 2023, urn:nasa:pds:maven.insitu.calibrated::36.0, PDSS, doi: 10.17189/ask6-5p89

Edberg, N. J. T., Nilsson, H., Williams, A. O., et al. 2010, GeoRL, 37, L0310720

Yoshida, N., Terada, N., Nakagawa, H., et al. 2021, JGRE, 126, e2021JE006926